\documentclass[conference]{IEEEtran}
\IEEEoverridecommandlockouts
\usepackage{textcomp}
\usepackage{xcolor}
\usepackage{cite}
\usepackage{latexsym}
\usepackage{graphicx}
\usepackage{amsfonts,amssymb,amsmath}
\usepackage{nccmath}
\usepackage{optidef}
\usepackage[hidelinks]{hyperref}
\usepackage{color,soul}
\usepackage{array}
\usepackage{algorithm}
\usepackage{algorithmic}
\usepackage[T1]{fontenc}
\usepackage[utf8]{inputenc}
\usepackage{fancyhdr}
\usepackage{lastpage}
\usepackage{subcaption}
\usepackage[abbreviations, automake]{glossaries-extra}
\makeglossaries 
\newabbreviation{OMA}{OMA}{orthogonal multiple access}
\newabbreviation{NOMA}{NOMA}{Non-orthogonal multiple access}
\newabbreviation{SINR}{{SINR}}{signal-to-interference-plus-noise ratio}
\newabbreviation{SIC}{{SIC}}{successive interference cancellation}
\newabbreviation{MSD}{{MSD}}{Minimum SINR  Difference}
\newabbreviation{CDF}{CDF}{cumulative distribution function}
\newabbreviation{NF}{NF}{Near-Far}
\newabbreviation{BS}{BS}{base station}
\newabbreviation{LR}{LR}{log rate}
\newacronym{N}{\ensuremath{N}}{Number of users connected to the base station}
\newacronym{GammaOMA}{\ensuremath{\gamma_i^\text{\tiny OMA}}}{SINR of the $i^{\text{th}}$ user in OMA system}
\newacronym{GammaNOMA}{\ensuremath{\gamma_i^\text{\tiny NOMA}}}{SINR of the $i^{\text{th}}$ user in NOMA system}
\newacronym{hi}{\ensuremath{h_i}}{Channel gain of the $i^{\text{th}}$ user}
\newacronym{N0}{\ensuremath{\sigma^2}}{Noise variance}
\newacronym{G}{\ensuremath{G}}{Number of users allocated with bandwidth}
\newacronym{I}{\ensuremath{I}}{Inter cell interference}
\newacronym{beta}{\ensuremath{\beta}}{Imperfect SIC parameter}
\newacronym{Pt}{\ensuremath{P_t}}{Available transmit power at the base station}
\newacronym{alphai}{\ensuremath{\alpha_i}}{Fraction of power allocated to the user $i$}
\newacronym{ROMA}{\ensuremath{R_i^\text{\tiny OMA}}}{Normalized downlink rate for a user $i$ in case of OMA system}
\newacronym{RNOMA}{\ensuremath{R_i^\text{\tiny NOMA}}}{Normalized downlink rate for a user $i$ in case of NOMA system}
\newacronym{Ffun}{\ensuremath{\chi_\text{\tiny G}}}{?}
\newacronym{Delta}{\ensuremath{\delta}}{?}
\newacronym{DeltaMSD}{\ensuremath{\Delta^{\tiny MSD}_{\tiny i-1, i}}}{?}
\newacronym{MUP}{MUC}{Multi-User Clustering}
\newacronym{AMUP}{AMUC}{Adaptive Multi-User Clustering}

\newacronym{NDRates}{\ensuremath{R_{s}^{\textmd{\tiny{NOMA}}}}}{Rate of the strong user in DL NOMA system}
\newacronym{NDRatew}{\ensuremath{R_{w}^{\textmd{\tiny{NOMA}}}}}{Rate of the weak user in DL NOMA system}

\newacronym{ROMAs}{\ensuremath{R_s^\text{\tiny OMA}}}{Normalized downlink rate for a user $i$ in case of OMA system}
\newacronym{ROMAw}{\ensuremath{R_w^\text{\tiny OMA}}}{Normalized downlink rate for a user $i$ in case of OMA system}

\usepackage{optidef}

\usepackage[english]{babel}
\usepackage{amsthm}

\DeclareMathAlphabet{\pazocal}{OMS}{zplm}{m}{n}

\def\BibTeX{{\rm B\kern-.05em{\sc i\kern-.025em b}\kern-.08em
    T\kern-.1667em\lower.7ex\hbox{E}\kern-.125emX}}
\begin{document}

\title{$\alpha$-Fairness User Pairing for Downlink NOMA Systems with Imperfect Successive Interference Cancellation  \\
}

\author{\IEEEauthorblockN{Nemalidinne Siva Mouni, Pavan Reddy M., Abhinav Kumar, and Prabhat K. Upadhyay}
{\centering \small{ Email: \{ee19resch11003, ee14resch11005\}@iith.ac.in, abhinavkumar@ee.iith.ac.in, pkupadhyay@iiti.ac.in}}
\vspace{-0.5cm}} 
\maketitle

\begin{abstract}
Non-orthogonal multiple access (NOMA) is considered as one of the predominant multiple access technique for the next-generation cellular networks. We consider a 2-user pair downlink NOMA system with imperfect successive interference cancellation (SIC). We consider bounds on the power allocation factors and then formulate the power allocation as an optimization problem to achieve {$\alpha$-Fairness} among the paired users. We show that {$\alpha$-Fairness} based power allocation factor coincides with lower bound on power allocation factor in case of perfect SIC and $\alpha > 2$. Further, as long as the proposed criterion is satisfied, it converges to the upper bound with increasing imperfection in SIC. Similarly, we show that, for $0<\alpha<1$, the optimal power allocation factor coincides with the derived lower bound on power allocation. Based on these observations, we then propose a low complexity sub-optimal algorithm. Through extensive simulations, we analyse the performance of the proposed algorithm and compare the performance against the state-of-the-art algorithms. We show that even though Near-Far based pairing achieves better fairness than the proposed algorithms, it fails to achieve rates equivalent to its orthogonal multiple access counterparts with increasing imperfections in SIC. Further, we show that the proposed optimal and sub-optimal algorithms achieve significant improvements in terms of fairness as compared to the state-of-the-art algorithms. 
\end{abstract}



%

\IEEEpeerreviewmaketitle 
\section{Introduction}\gls{NOMA} is a promising technology  that has capability to fulfill diverse requirements of next generation multiple access (NGMA)~\cite{Ding}. It achieves significant data rates for a large number of users and efficient utilization of the spectrum, thus, becoming a key spectrally efficient technology. 
There exists different approaches for \gls{NOMA}, which can broadly be classified into two categories as Power-Domain \gls{NOMA} and Code-Domain \gls{NOMA}. In Power Domain-\gls{NOMA}, the base station allocates different power allocation factors to the multiplexed users, whereas, in the Code-Domain \gls{NOMA}, the users are multiplexed in the code domain~\cite{dai}.
The main idea behind PD-\gls{NOMA} is allowing multiple users to access the same resource block but with varying power levels. The two key principles of \gls{NOMA} are superposition coding and \gls{SIC}. The transmitter superimposes the information related to multiple users and transmits in a single resource block (same space-time-frequency resource), i.e., the transmitter performs superposition coding. At the higher channel gain user, superposed information needs to be decoded and \gls{SIC} serves this purpose~\cite{Benj}, whereas, the user with lower channel gain decodes its information by considering the information related to higher channel gain user as noise.

\gls{NOMA} can combine with diverse technologies and help in increasing the capacity of the system, providing user fairness and achieving spectral efficiency~\cite{MVaezi}. Most of the researchers from academia, industry as well as policy making are actively investigating \gls{NOMA}. However, there are many design and implementation issues that needs to be addressed. User pairing and power allocation strategies have always been crucial to achieve better \gls{NOMA} performance. Given a \gls{NOMA} user pair, the authors in~\cite{Benj} validated that the individual user rates exceed its corresponding \gls{OMA} rates. This is always true for perfect \gls{SIC} case, wherein, we assume that the higher channel gain user decodes its information perfectly. However, in a practical scenario, there is always some imperfection associated with \gls{SIC} due to hardware impairments or implementation issues such as complexity scaling and error propogation~\cite{SIC}. This phenomenon impacts the achievable rate of the user with higher channel gain in a \gls{NOMA} pair, directly affecting the overall throughput of the system. Hence, the network operators have to consider the impact of imperfections in \gls{SIC} while performing scheduling and power allocation to realize the true benefits of \gls{NOMA}. 

Some works in the existing literature have considered the impact of imperfection in \gls{SIC}~\cite{SIC,Mouni}. However, this is the first work that consider the imperfection in \gls{SIC} while performing $\alpha$-~Fairness based power allocation for a 2-user pair in the downlink \gls{NOMA} system. In view of the aforementioned details, we present the following key contributions in this paper.

\begin{itemize}

\item [1.] We present a detailed analysis on $\alpha$-Fairness among the 2-user pair in a practical \gls{NOMA} system with imperfect SIC. 

\item [2.] We formulate maximizing the utility function as an optimization problem for the downlink \gls{NOMA} system in the presence of imperfect \gls{SIC}.

\item [3.] We propose a low-complexity sub-optimal algorithm for $\alpha$-Fairness-based power allocation, performing similar to optimal algorithm based on the optimization problem formulated.

\item [4.] We perform extensive numerical evaluations and show that the proposed algorithms significantly outperform the state-of-the-art algorithms.

\end{itemize}

The organization of the paper is as follows. In Section~\ref{sec:RelatedWork}, we present relevant related works in the literature. The system model is explained in Section III along with brief introduction on the upper and lower bounds on power allocation factors. The $\alpha$-Fairness based optimization problem is formulated in Section~IV, with explanation of the optimal and sub-optimal algorithm. Numerical results are presented in Section~V with detailed explanation of plots generated. Section~VI comprises of some concluding remarks and the scope of future works.

\section{Related Works}
\label{sec:RelatedWork}
In this section, we will discuss a few relevant fairness based formulations and proposed schemes for \gls{NOMA} systems. Most of the state-of-the-art techniques proposed for user pairing and power allocation focus on improving the overall system throughput at the expense of fairness among the paired users. Moreover, the existing works which considered fairness such as~\cite{woSIC1,PF1, PF2}, have not analyzed or studied the impact of imperfection in \gls{SIC}. In \cite{woSIC1}, authors have implemented a lossless \gls{NOMA} without \gls{SIC} and proved the conditional achievable sum-rate given channel gain realizations in the conventional \gls{NOMA} scheme without \gls{SIC}, without loss, over the power allocation range of user-fairness. The authors in \cite{PF1} have proposed a joint user pairing and power allocation algorithm in the \gls{NOMA} uplink communication systems, aiming at improving the proportional fairness of the users. The authors in \cite{PF2} have proposed a new resource allocation scheme for a downlink OFDMA-based \gls{NOMA} system to maximize the sum capacity performance under a general proportional user fairness constraint. A spectrum resource and power allocation algorithm with adaptive proportional fair user pairing has been proposed to convert the formulated optimization problem into user pairing, sub-channel, and power allocation in \cite{PF3}. 

Some authors have also considered max-min fairness apart from the proportional fairness based user pairing and power allocation schemes for \gls{NOMA} as discussed in \cite{PF1, PF2, PF3}. Authors in \cite{MM1} proposed optimal power allocation rules in a variety of problem settings: 1) max-min fairness, 2) sum-rate maximization with a minimum data rate constraint, and 3) weighted sum-rate maximization for improving data rates of cell-edge users. The resource allocation problem for \gls{NOMA}-enabled V2X communications has been investigated in \cite{MM2}, where weighted max-min rate fairness is applied to achieve user fairness and the different requirements of cellular users.

In~\cite{OPT}, $\alpha$-Fairness based power allocation schemes for sum throughput and ergodic rate maximization problems in a downlink \gls{NOMA} system have been analyzed. The resource allocation fairness of the \gls{NOMA} and \gls{OMA} schemes have been investigated in \cite{FC1}, wherein the fundamental reason of \gls{NOMA} being more fair than OMA in asymmetric multiuser channels has been analyzed. The authors in \cite{F_SR} have maximized the average sum rate ensuring a minimum average rate for each user by optimally adapting the power and rate allocations to varying fading states. However, none of the works have considered the fairness among the paired users and the trade-off with imperfect \gls{SIC} to achieve such fairness for a downlink \gls{NOMA} system.

\section{System Model}

\begin{figure}
  \includegraphics[scale=.3]{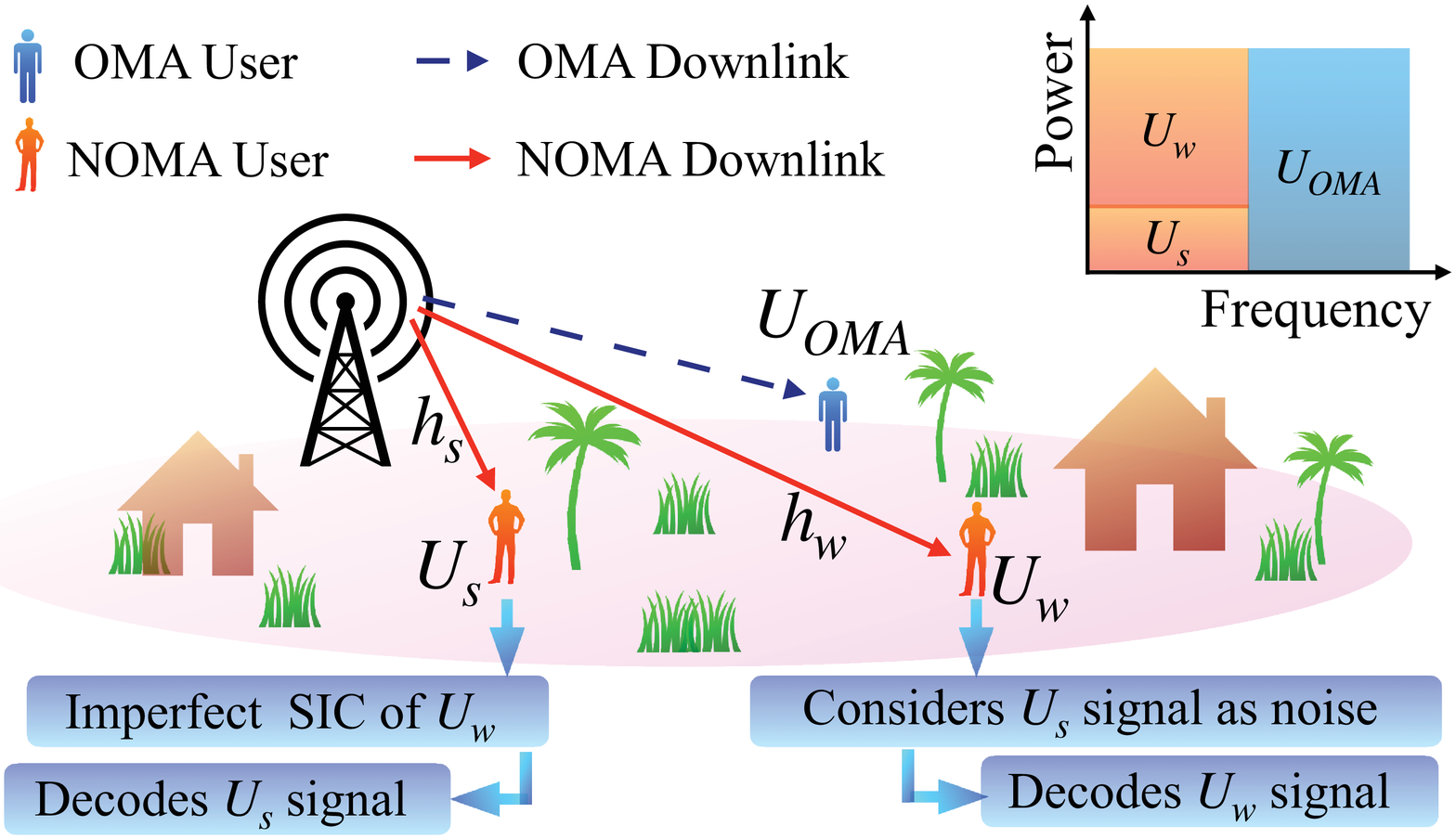}
  \caption{System Model.}
   \label{fig:SM}
\end{figure}

Without loss of generality, we consider a 5G cellular network as shown in Fig. \ref{fig:SM} and focus on the downlink \gls{NOMA} 2-user pairing scheme. In a typical \gls{OMA} system, the downlink \gls{SINR} from the transmitter (i.e., the base station) to a receiver (user $u$), on a subchannel, is formulated as

\begin{equation}
\gamma_{u} = P_t \frac{|h_u|^2}{\sigma^{2} + I},
 \label{eqn:SINRo_OMA}\end{equation}
where $P_t$ is the transmitted power, $h_u$ is the channel gain of user $u$, \color{black} $\sigma^{2}$ is the noise power, and $I$ is the interference received on the subchannel allocated to the user $u$ from nearby base stations. The normalized downlink rate for such a user in an \gls{OMA} based \gls{LR} model is 

\begin{equation}
 R_{u}^{\textmd{\tiny{\gls{OMA}}}} =   \frac{1}{2}\log_{2}\left (1 + \gamma_{u} \right),
 \label{eqn:SINR_OMA}
\end{equation}
\color{black} where the factor ${1}/{2}$ is considered owing to the loss in multiplexing in \gls{OMA} system.

\subsection{\gls{NOMA} Rate Formulation}
Consider a strong user $U_s$ and weak user $U_w$ \gls{NOMA} pair, such that the relation between their channel gains from a particular base station is  $|h_s|^2 > |h_w|^2$. For the \gls{NOMA} \gls{LR} formulation, let us assume the fraction of power allocated to user $U_s$ is $\delta_s$ and $U_w$ is $1 - \delta_s$. Let $\beta \in [0,1]$ be the imperfection in \gls{SIC} associated with the strong user resulting from implementation issues such as complexity scaling and error propagation~\cite{SIC}. The \gls{SINR} of the \gls{NOMA} user pair, i.e., $\hat{\gamma}_{s}$ and $\hat{\gamma}_{w}$ can be, respectively, expressed as~\cite{SIC} 
\begin{align}
    \hat{\gamma}_{s} &=  \frac{\delta_sP_t |h_s|^2}{ \sigma^{2} + I + \beta(1-\delta_s)P_t |h_s|^2 } \nonumber \, \mbox{ and} \\
    \hat{\gamma}_{w} &= \frac {(1-\delta_s)P_t |h_w|^2}{\sigma^{2} + I + \delta_sP_t |h_w|^2},
    \label{eqn:SINR_NOMA}
\end{align} 
\color{black}
where $\beta = 0$ implies perfect \gls{SIC} case, i.e., the strong user is completely able to remove the interference from the weak user. Following similar formulations as in~\cite{Mouni}, we reformulate  (\ref{eqn:SINR_NOMA}) using (\ref{eqn:SINR_OMA}) as follows

\begin{gather}
\hat{\gamma}_{s} = \frac{\delta_s\gamma_s}{ 1 + \beta(1-\delta_s)
\gamma_s} \mbox{ and} \hspace{0.2 cm}
  \hat{\gamma}_{w} =  \frac {(1-\delta_s)\gamma_w}{1 + \delta_s\gamma_w}.
\end{gather}
Further, the \gls{NOMA} rates of the user pair, i.e., $R_{s}^{\textmd{\tiny{\gls{NOMA}}}}$ and $R_{w}^{\textmd{\tiny{\gls{NOMA}}}}$, are respectively given by
\color{black}
\begin{gather}
   R_{s}^{\textmd{\tiny{\gls{NOMA}}}} =  \log_{2} \left (1 + \hat{\gamma}_{s} \right)  \mbox{ and} \hspace{0.2 cm}
    R_{w}^{\textmd{\tiny{\gls{NOMA}}}} =  \log_{2} \left (1 + \hat{\gamma}_{w} \right).
    \label{eqn:Rates_NOMA}
\end{gather}



\subsection{Bounds on Power Allocation Factor}
We consider the upper and lower bounds on the power allocation factor $\delta_s$ from our previous work~\cite{Mouni}. As shown in~\cite{Mouni}, solving $R_{w}^{\textmd{\tiny{\gls{NOMA}}}} > R_{w}^{\textmd{\tiny{OMA}}}$ results in (\ref{eqn:W_alpha}), which ensures weak user NOMA rate is greater than its OMA rate. Similarly, solving $R_{s}^{\textmd{\tiny{\gls{NOMA}}}} > R_{s}^{\textmd{\tiny{OMA}}}$ results in $\delta_{lb}$ (\ref{eqn:S_alpha}), such that strong user \gls{NOMA} rate is greater than its OMA counterpart.

\begin{equation}
\delta_{ub} = \frac{1}{\gamma_w}\big(\sqrt{1 + \gamma_w} - 1 \big) ,
 \label{eqn:W_alpha}
\end{equation}

\begin{equation}
\delta_{lb} = \frac{(1 + \beta \gamma_s)(\sqrt{1 + \gamma_s} -  1)}{ \gamma_s(1 + \beta \sqrt{1+\gamma_s} - \beta)}.
 \label{eqn:S_alpha}
\end{equation}
Considering individual rates with \gls{NOMA} to be better than the corresponding OMA rates as shown in \eqref{eqn:Rates_NOMA} and solving for achievable sum-rate with \gls{NOMA} to be better than achievable sum-rate with OMA,  the minimum \gls{SINR} difference (MSD) criterion for two users to be paired is formulated as~\cite{Mouni}  
\begin{equation}
\Delta^{*}_{MSD} = \gamma_s - \frac{(\sqrt{1 + \gamma_w} - 1)(\sqrt{1 + \gamma_s}\sqrt{1 + \gamma_w} + 1)}{\sqrt{1 + \gamma_w}} .
 \label{eqn:MSD}
\end{equation}
For a given pair of two users, if $\gamma_s - \gamma_w > \Delta^{*}_{MSD}$, then we will be able to find a $\delta_s$ between $\delta_{lb}$ and $\delta_{ub}$ such that individual \gls{NOMA} rates exceed their OMA counterparts, provided $\beta_s<\beta^*$. Here, $\beta$ is the imperfection in \gls{SIC} and $\beta^*$ is the upper bound on the same, formulated as  
\begin{equation}
\beta < \frac{\gamma_w - \gamma_s  + \gamma_s\sqrt{1+\gamma_w} - \gamma_w\sqrt{1 + \gamma_s} }{\gamma_s(\sqrt{1 + \gamma_s} - 1)(\gamma_w  - \sqrt{1 + \gamma_w} +1)} \triangleq \beta^{*}.
\label{eqn:beta}
\end{equation}
 Readers unfamiliar with the MSD criterion and the bounds on the imperfect \gls{SIC} are suggested to read our previous work which has detailed explanation for the same ~\cite{Mouni}.

\subsection{$\alpha$-Fair Scheduler}
The utility function for an $\alpha$-Fair scheduler with the variable $x$ is expressed as~\cite{Akhil}
\begin{equation}
U_{\alpha}(x) = 
     \begin{cases}
      {\frac{x^{1 - \alpha}}{1 -\alpha}}&\quad\text{if }  \alpha > 0, \alpha \neq 1,\\
      \log(x) &\quad\text{if } \alpha = 1.
      \label{eqn:U_alpha}
    \end{cases}
\end{equation}
The performance metric to measure the system performance of \gls{NOMA} user rates, i.e., $\alpha -$ Fair throughput is~\cite{Akhil}

\begin{align}
T_{\alpha}(x) = 
    \begin{cases}
        (\frac{1}{2}(R_{s}^{\textmd{\tiny{\gls{NOMA}}}})^{1 - \alpha}+ (R_{w}^{\textmd{\tiny{\gls{NOMA}}}})^{1 - \alpha})^{(\frac{1}{1-\alpha})} &\\
     &\hspace{-2cm}\quad\text{if }  \alpha > 0, \alpha \neq 1,\\
     (R_{s}^{\textmd{\tiny{\gls{NOMA}}}} R_{w}^{\textmd{\tiny{\gls{NOMA}}}})^{\frac{1}{2}} &\hspace{-2cm}\quad\text{if } \alpha = 1.
  \label{eqn:T_alpha}
    \end{cases}
\end{align}
Next, we present the proposed algorithms.

\section{Proposed Algorithms}
In this section, we initially formulate the power allocation for the paired users as an optimization problem and then present a sub-optimal algorithm that achieves close to optimal performance.


\subsection{Optimal Algorithm}
Given the rates of strong and weak user as in (\ref{eqn:SINR_NOMA}), we maximize the utility function presented in \eqref{eqn:U_alpha} while ensuring a minimum of \gls{OMA} rates for each paired user.
Thus, we formulate the optimization problem as follows

\begin{align}
\mathbf{P1}: \max_{\delta_s} \quad &  \left (  U_{\alpha}(\gls{NDRates}) +  U_{\alpha}(\gls{NDRatew}) \right ) \label{eqn:OBJ} \\
\textrm{s.t.} \quad & \gls{ROMAs} < \gls{NDRates}, \label{eqn:C_S}\\
\quad & \gls{ROMAw} < \gls{NDRatew}, \label{eqn:C_W} \\
   & \delta_s < \delta_{ub}, \label{eqn:C_ub}\\
   & \delta_s > \delta_{lb}, \label{eqn:C_lb}\\
   &  0 < \delta_s, \delta_w < 1,  \label{eqn:C_bs}\\
  & 0\leq \delta_s + \delta_{w} \leq  1, \label{eqn:C_SA}
\end{align}
where (\ref{eqn:OBJ}) is the overall objective function for maximizing the $\alpha$-fairness among the \gls{NOMA} pair under consideration over $\delta_s$. The constraints in (\ref{eqn:C_S}) and (\ref{eqn:C_W}) ensure that the strong and weak user \gls{NOMA} rates exceed their OMA counterparts. We need to pick $\delta_s$ between the upper and lower bounds as in \cite{Mouni}, and hence, we consider (\ref{eqn:C_ub}) and (\ref{eqn:C_lb}).
The constraint in (\ref{eqn:C_bs}) ensures that the fraction of power allocated to strong and weak user lie between $0$ and $1$. Moreover, in downlink \gls{NOMA}, their sum is $1$, thus, we have (\ref{eqn:C_SA}).
The given problem $\mathbf{P1}$ in (\ref{eqn:OBJ}) is a non linear programming problem (NLP), which is difficult to solve with first order differentiation. Hence, next, we present a low-complexity sub-optimal algorithm that can be realized in practical implementation and achieves close to optimal performance.


\subsection{Sub-optimal Algorithm}
\begin{algorithm}[t]
 INPUTS : $ \hat{\gamma}_{s}, \hat{\gamma}_{w}, \beta_s, \alpha, \tau $ \\
 OUTPUTS : $\delta_s$ \\
 \textbf{Optimal Algorithm:} \\
 1. \textbf{if} $ \hat{\gamma}_{s} -  \hat{\gamma}_{w} > \Delta^{*}_{MSD}$ and $\beta_s < \beta^*$ \newline
 2.\hspace{0.3 cm} Solve \textbf{P1} to get $\delta_s$ \newline
 3.\hspace{0.3 cm} Pair $ \hat{\gamma}_{s}, \hat{\gamma}_{w}$ using the resultant $\delta_s$ from \textbf{P1}\newline
 4. \textbf{else}\newline
 5.\hspace{0.3 cm} Consider $\hat{\gamma}_{s}, \hat{\gamma}_{w}$ as \gls{OMA} users.\newline   
 6. \textbf{end}\newline
 \newline
 \textbf{Sub-Optimal Algorithm:}\\
 1. \textbf{if} $ \hat{\gamma}_{s} -  \hat{\gamma}_{w} > \Delta^{*}_{MSD}$ \newline
 2.\hspace{0.3 cm} \textbf{if} ( $\beta_s / \beta^* < \tau $)\newline
 3.\hspace{0.75 cm} \textbf{if} ( $\alpha >1 $)\newline
 4.\hspace{1.05 cm} $\delta_s = \delta_{lb}$\newline
 5.\hspace{0.75 cm} \textbf{else if} ($0 < \alpha \leq 1$ ) \newline
 6.\hspace{1.05 cm} $\delta_s = \delta_{ub}$ \newline
 7.\hspace{0.75 cm} \textbf{end}  \newline
 8.\hspace{0.3 cm} \textbf{else}\newline
 9.\hspace{0.75 cm} \textbf{if} ( $\alpha >1 $)\newline
 10.\hspace{1.05 cm} $\delta_s = \delta_{ub}$\newline
 11.\hspace{0.62 cm} \textbf{else if} ($0 < \alpha \leq 1$ ) \newline
 12.\hspace{0.95 cm} $\delta_s = \delta_{ub}$ \newline
 13.\hspace{0.65 cm} \textbf{end}  \newline
 14.\hspace{0.29 cm} \textbf{end} \newline
 15. \textbf{else} \newline
 16.\hspace{0.3 cm}  Consider $\hat{\gamma}_{s}, \hat{\gamma}_{w}$ as \gls{OMA} users.\newline    
 17. \textbf{end}\newline
 \caption{\!Proposed \! Algorithms.}
\end{algorithm}



Considering the strong user rate is greater than the weak user rate and $\alpha<1$, because of the positive exponentials in the utility function, the optimization formulation in \eqref{eqn:OBJ} is significantly dependent on  $R_{s}^{\textmd{\tiny{\gls{NOMA}}}}$ in perfect \gls{SIC} case. Further, a smaller increase in $\delta_s$ significantly increases the  $R_{s}^{\textmd{\tiny{\gls{NOMA}}}}$ and slightly decreases the  $R_{w}^{\textmd{\tiny{\gls{NOMA}}}}$. Hence, larger the value of $\delta_s$, greater will be the value of the objective function in \eqref{eqn:OBJ}. Thus, for $0<\alpha\leq 1$, we choose $\delta_s=\delta_{ub}$. Similarly, for $\alpha>1$, because of the negative exponentials in the utility functions, the optimization problem formulated in \eqref{eqn:OBJ} is significantly dependent on $R_{w}^{\textmd{\tiny{\gls{NOMA}}}}$, and hence, the smaller the value of $\delta_s$, the greater will be the value of the objective function in \eqref{eqn:OBJ}. Thus, for $\alpha>1$, we choose $\delta_s=\delta_{lb}$. However, with an increase in the $\beta$ value, the $R_{s}^{\textmd{\tiny{\gls{NOMA}}}}$ value decreases. Hence, for larger $\beta$, to ensure minimum \gls{OMA} rates for the strong user, we need to choose $\delta_s=\delta_{ub}$. Note that we consider the users for \gls{NOMA} pairing only when they meet the minimum \gls{SINR} criterion presented in \eqref{eqn:MSD}. Otherwise, we consider them for \gls{OMA}.

For a given \gls{NOMA} pair, we denote the worst case imperfection in \gls{SIC} possible as $\beta_s$ and the maximum permissible imperfection in \gls{SIC} as $\beta^*$ (obtained from \eqref{eqn:beta}). In the sub-optimal algorithm, we consider $\beta_s$ to be a smaller value when it satisfies $(\beta_s/\beta^*)< \tau $. Otherwise, we consider $\beta_s$ to be a larger value and closer to $\beta^*$.  Then, we allocate the power to the users based on the earlier explanation. We have summarized the entire procedure for optimal and sub-optimal power allocation in Algorithm~1.
\subsection{Baseline Algorithms considered for Evaluation}
\subsubsection{\gls{NF}} 
For comparison with the proposed $\alpha-$Fairness-based power allocation, we consider the \gls{NF} algorithm presented in \cite{NearFar1}. The users are initially sorted in decreasing order of their channel gains. Then, two users are picked sequentially one from the beginning and the other from the end of the sorted user list and paired together for \gls{NOMA} transmission. The algorithm achieves significant improvement in the network throughput because of the sorting. However, the algorithm does not consider the practical imperfections in \gls{SIC} while pairing users, and hence, the performance degrades with the increase in the imperfections in \gls{SIC}.
\subsubsection{\gls{OMA}} For the same set of users, we evaluate the network performance with \gls{OMA}. For this scenario, we consider the formulations presented in \eqref{eqn:SINRo_OMA}-\eqref{eqn:SINR_OMA}.
\subsubsection{Derived bounds} We also evaluate the network performance by considering the power allocation factors equal to the derived upper and lower bounds formulated in \eqref{eqn:W_alpha}-\eqref{eqn:S_alpha}. Further, in this evaluation, we also consider the MSD formulated in \eqref{eqn:MSD} and the derived bounds on the imperfection in \gls{SIC} presented in~\eqref{eqn:beta}.
Next, we present the detailed simulation set up and discuss the numerical results.

\section{Results and Discussion}

%
\label{sec:Results}
\begin{figure}
  \centering
  \includegraphics[scale=.4]{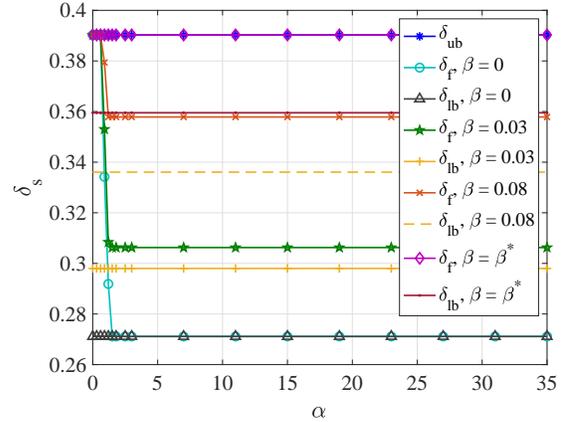}
  \caption{ Comparison of $\delta_s$ for varying $\alpha$.}
  \label{fig:del2}
\end{figure}

\begin{figure}
  \centering
    \includegraphics[scale=.4]{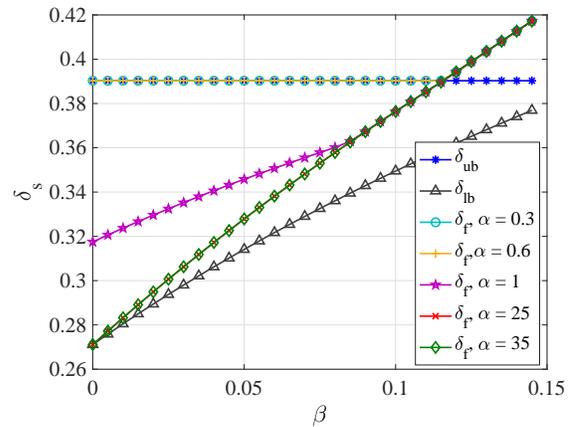}
  \caption{Comparison of $\delta_s$ for varying $\beta$.}
  \label{fig:del3}
\end{figure}

\begin{figure*}
\begin{subfigure}{0.33\textwidth}
  \centering
    \includegraphics[scale=.45]{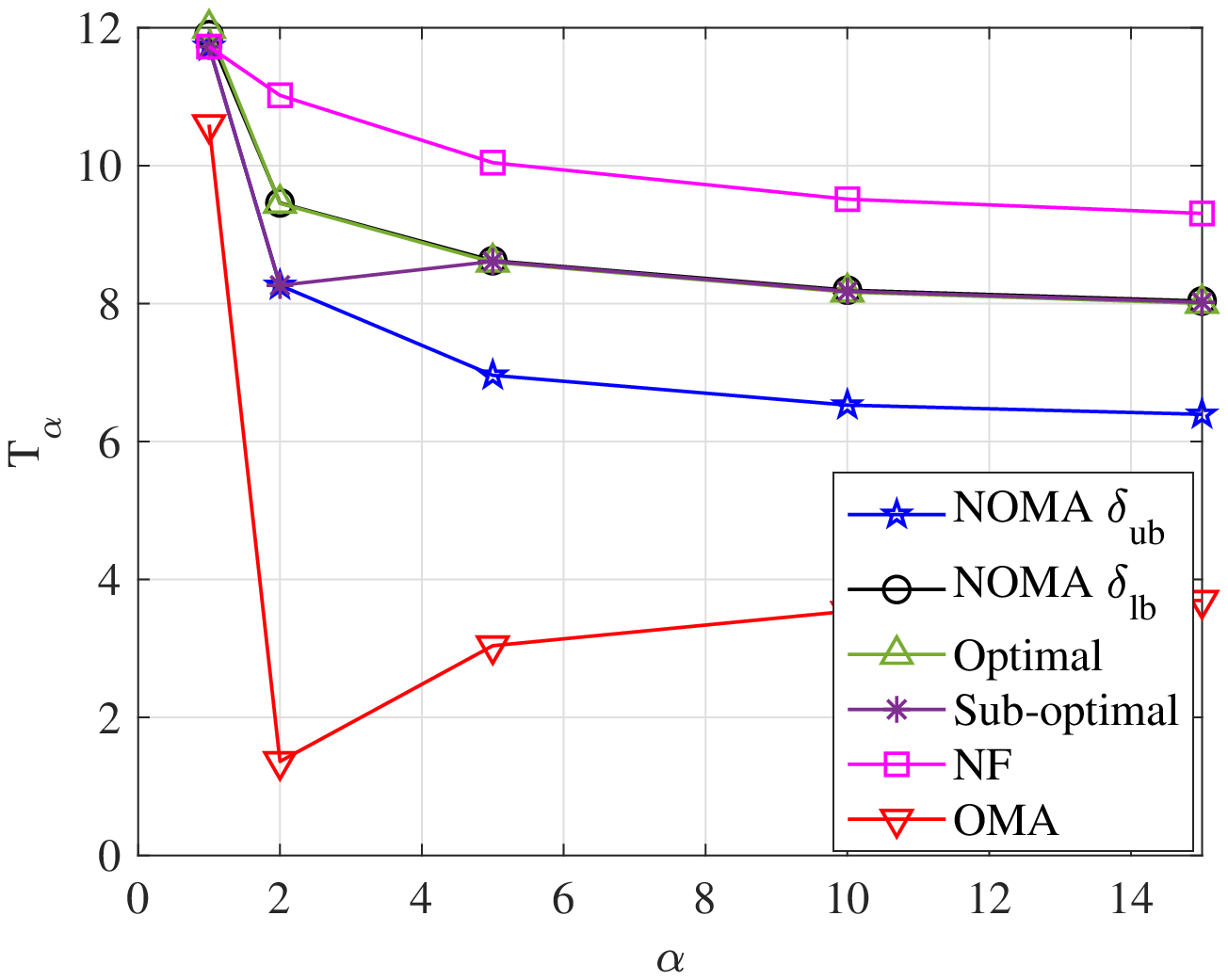}
  \caption{$T_{\alpha}$ vs $\alpha$ with $\beta$ = 0.01.}
  \label{fig:TA1}
\end{subfigure}
\begin{subfigure}{0.33\textwidth}
    \centering
  \includegraphics[scale=.45]{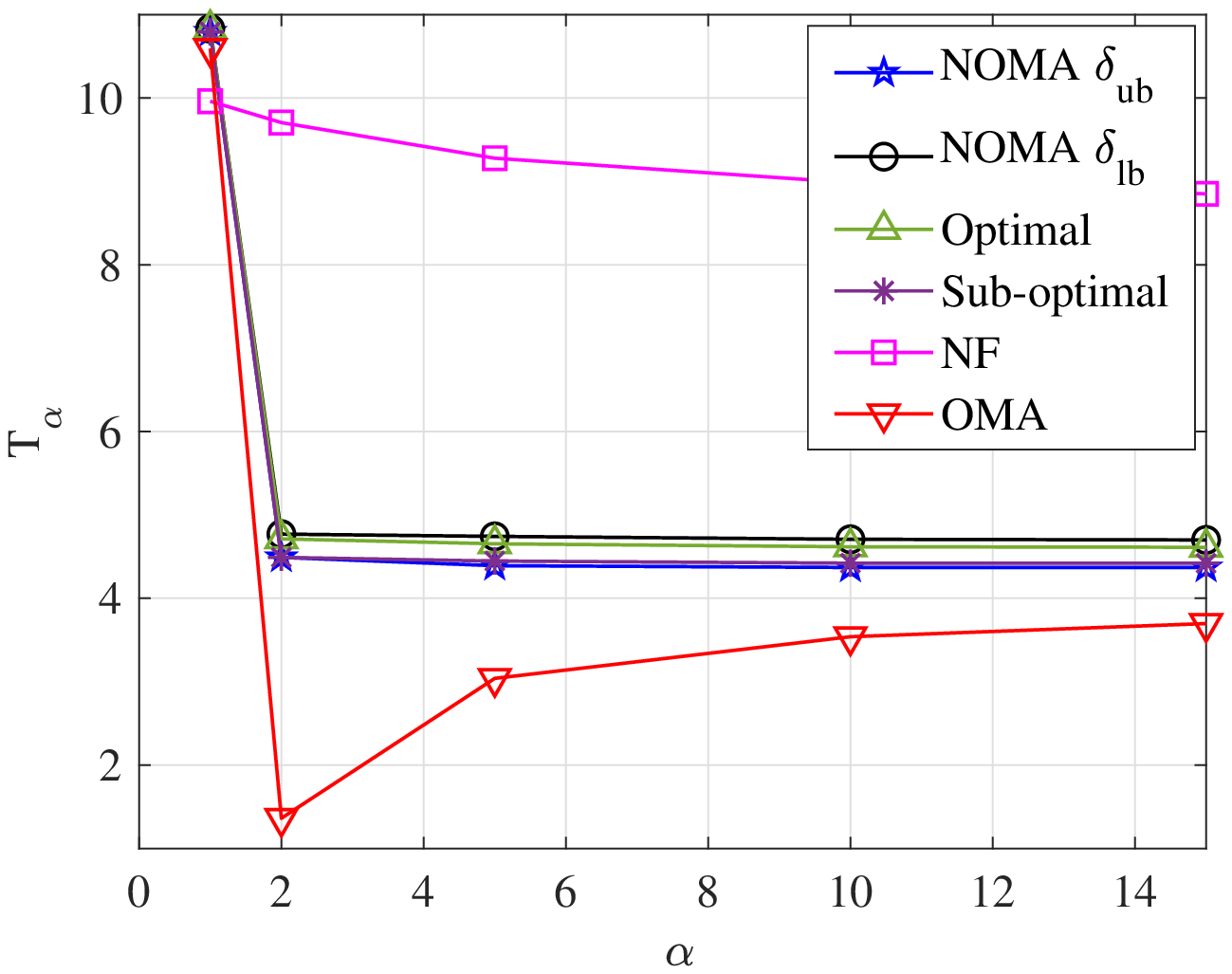}
  \caption{$T_{\alpha}$ vs $\alpha$ with $\beta$ = 0.06.}
  \label{fig:TA6}
\end{subfigure}
  \begin{subfigure}{0.33\textwidth}
   \centering
   \includegraphics[scale=.45]{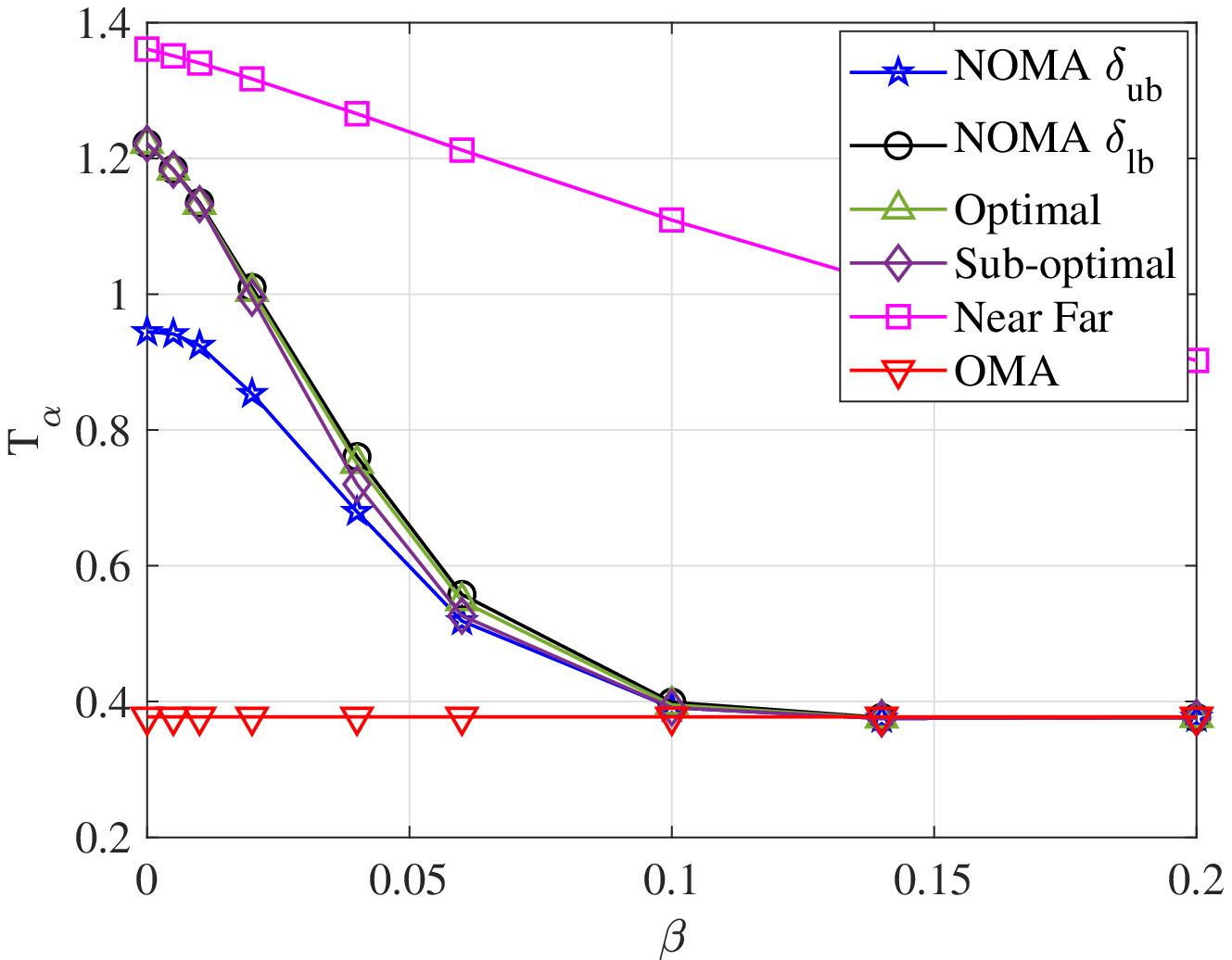}
  \caption{$T_{\alpha}$ vs $\beta$ with  $\alpha$ = 1.}
  \label{fig:TB1}
\end{subfigure}
  \caption{Comparison of $T_{\alpha}$ for varying $\alpha$ and $\beta$.}
\end{figure*}
We consider Poisson point distributed base stations and user positions with densities $25$ BS/km$^{2}$ and $120$ users/km$^{2}$, respectively, as in~\cite{Nemati}. We perform Monte Carlo simulations for an urban cellular environment pathloss model~\cite{std} considering Rayleigh fading model (scaling parameter $ = 1$) and omni-directional antennas. The user association to a particular BS is based on the maximum \gls{SINR} received by all the BSs. We study the variation in the fraction of power allocated for different values of $\alpha$ by considering both perfect and imperfect \gls{SIC} cases. We compare the performance of optimal and sub-optimal algorithms (assuming $\tau =0.5$) presented in \textbf{P1} with other MSD based pairing strategies: $\delta_s = \delta_{ub}$, $\delta_s = \delta_{lb}$, \gls{NF} pairing scheme with $\delta_s = \delta_{ub}$, and \gls{OMA}.

Fig. \ref{fig:del2} shows the variation in $\delta_s$ for varying $\alpha$, considering different values of $\beta$, i.e., $\beta = [0, 0.03, 0.08, \beta^*]$ for $\gamma_s = 9$~dB, $\gamma_w = 2 $~dB. The resultant fairness based power allocation factor from the optimization problem, i.e., $\delta_f$ is equivalent to lower bound in case of perfect \gls{SIC} and higher $\alpha$. This $\delta_f$ tends towards the upper bound with increase in imperfection in \gls{SIC} as observed for $\beta = [0.03, 0.08]$ in Fig. \ref{fig:del2}. Further, it is equivalent to upper bound, i.e., $\delta_{ub}$ when $\beta = \beta^{*}$  for all values of $\alpha$. Thus, we note the following important remarks observed from Fig. \ref{fig:del2} which are aligned with the formulations in our proposed sub-optimal algorithm:   
\\
\textit{Remark 1:}
Given a perfect \gls{SIC} \gls{NOMA} system i.e., $\beta = 0$, the fairness based power allocation $\delta_{f}$ is equivalent to the lower bound $\delta_{lb}$ $\forall$ $\alpha > 2$. 
\\
\textit{Remark 2:}
For $ 0 < \alpha < 1$, the fairness based power allocation $\delta_{f}$ is equivalent to the upper bound $\delta_{ub}$, $\forall$ $\beta < \beta^{*}$. 

Fig. \ref{fig:del3} illustrates the variation of $\delta_s$ with imperfection in \gls{SIC} for different values of $\alpha$, i.e., $\alpha = [0.3, 0.6, 1, 25, 35]$. For $0 < \alpha  < 1$, the $\alpha$-Fairness based power allocation is equivalent to the upper bound, for $\beta < \beta^{*}$. Further, for $\beta =0$, we observe that the $\alpha$-Fairness based power allocation is equivalent to lower bound $\delta_{lb}$ for any $\alpha > 1$. However, for higher values of $\alpha$ and $\beta < \beta^*$, the resultant $\delta_s$ from the optimal algorithm in \textbf{P1} always lies between the upper and lower bounds on $\delta_s$. Further, with increase in $\beta$, the value moves away from lower bound and converges with the upper bound as shown in Fig. \ref{fig:del3}. This illustration gives us another important remark which is aligned with our formulation in the sub-optimal algorithm:      
\\
\textit{Remark 3:}
Given an imperfect \gls{SIC} \gls{NOMA} system where $\beta = \beta^*$, the fairness based power allocation $\delta_{f}$ is equivalent to the upper bound $\delta_{ub}$, for any $\alpha$.

\begin{figure*}
\begin{subfigure}{0.33\textwidth}    \centering
  \includegraphics[scale=.38]{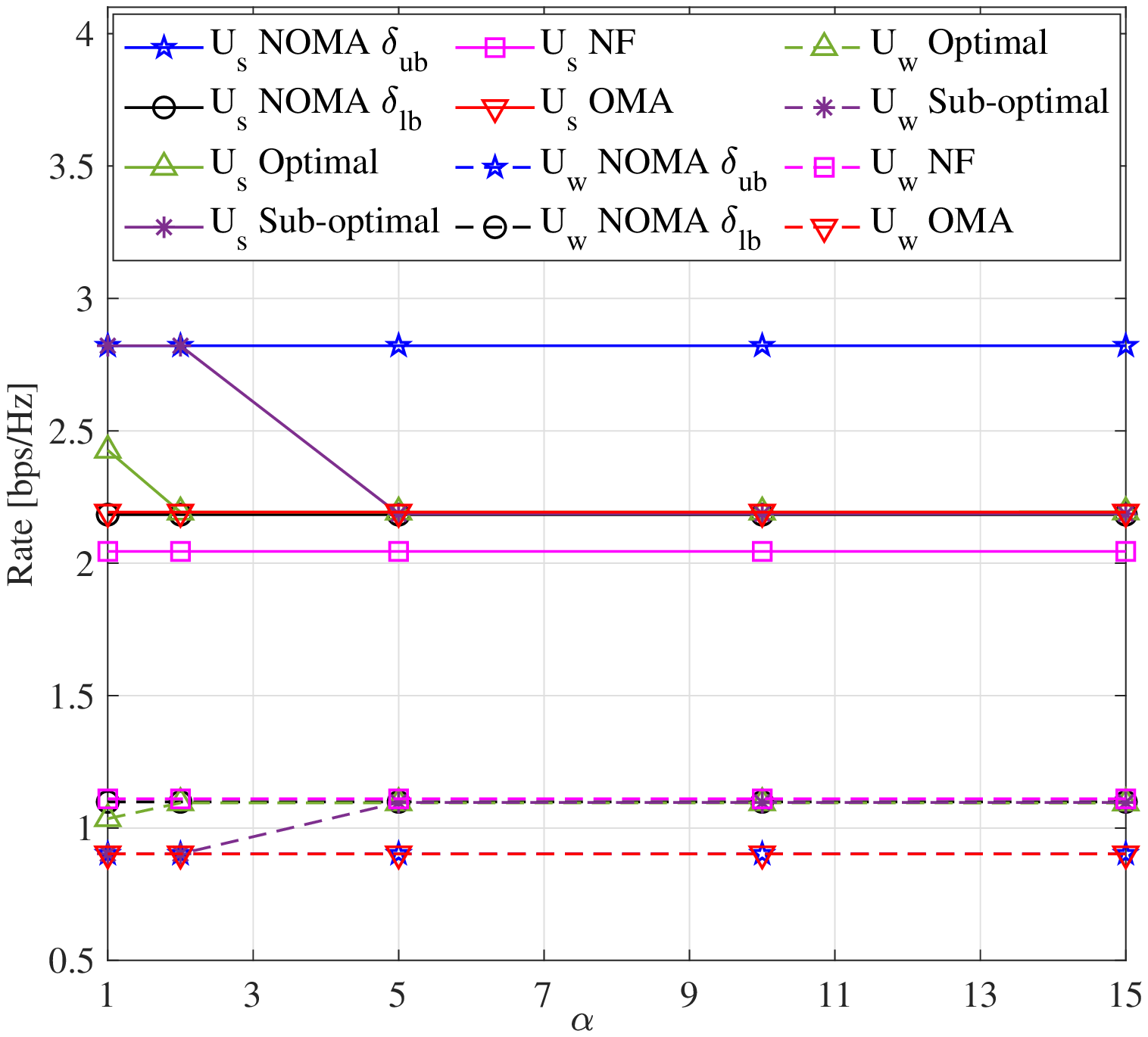}
  \caption{MUR vs $\alpha$ with $\beta$ = 0.01.}
  \label{fig:MA1}
\end{subfigure}
  \begin{subfigure}{0.33\textwidth}
    \centering
  \includegraphics[scale=.38]{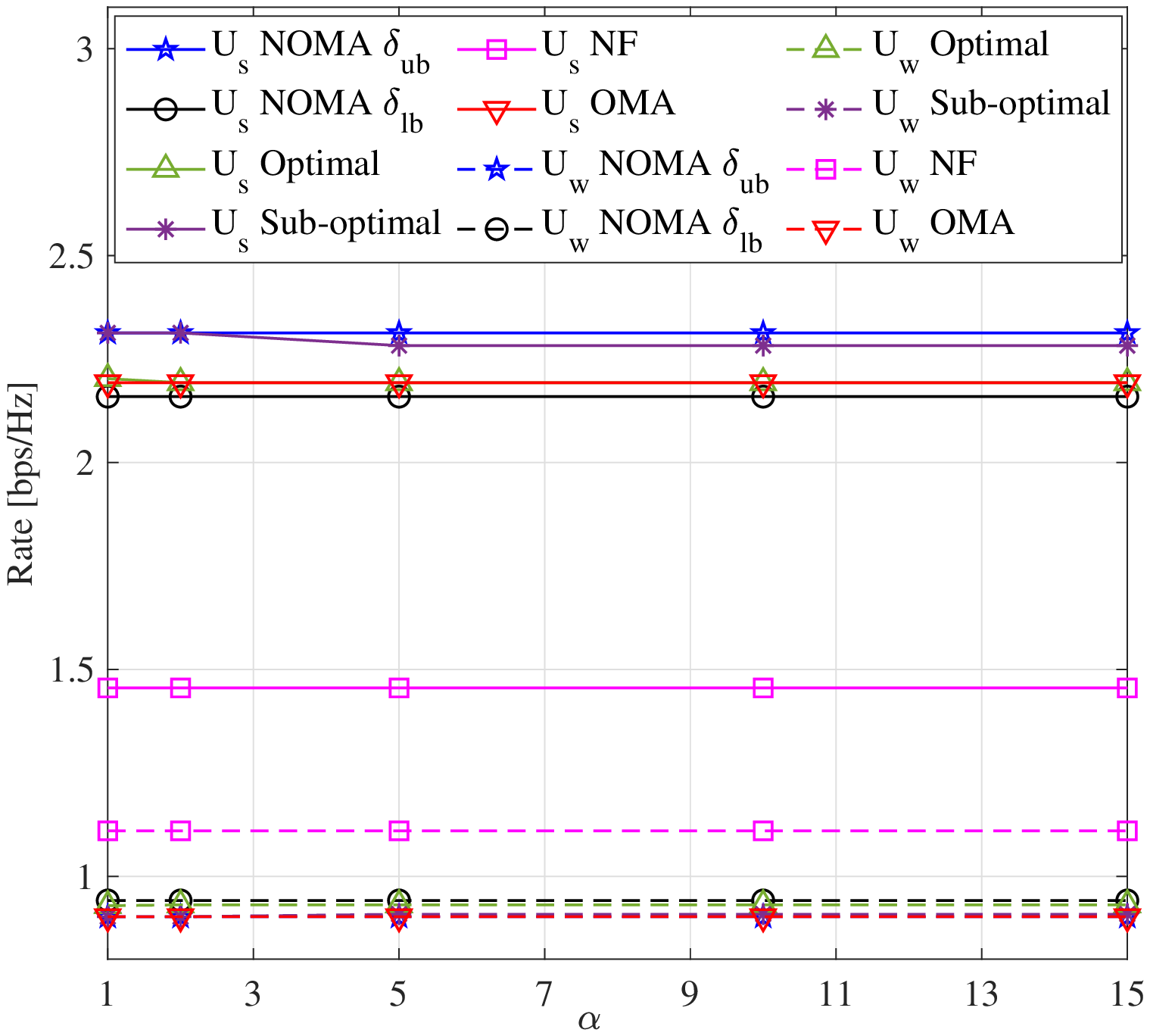}
  \caption{ MUR vs $\alpha$ with $\beta$ = 0.06.}
  \label{fig:MA6}
\end{subfigure}
  \begin{subfigure}{0.33\textwidth}
    \centering
  \includegraphics[scale=.38]{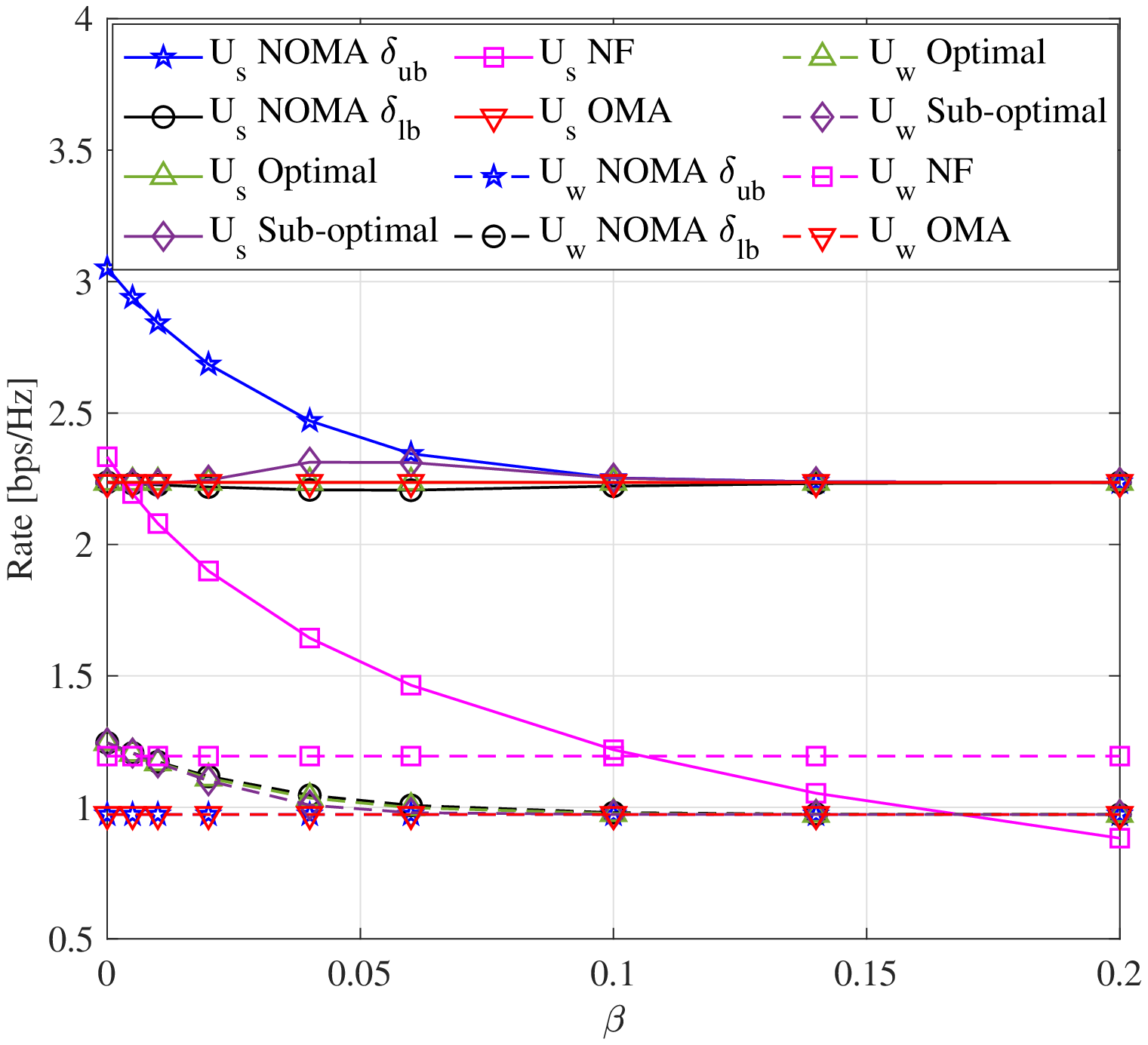}
  \caption{MUR vs $\beta$ with $\alpha$ = 1.}
  \label{fig:MAA1}
\end{subfigure}
\caption{Comparison of MUR for varying  $\alpha$ and $\beta$.}
\end{figure*}


Fig. \ref{fig:TA1} and Fig. \ref{fig:TA6} illustrate the variation of $T_{\alpha}$ with respect to different values of $\alpha$, i.e., [1, 2, 5, 20] for $\beta = 0.01$ and $\beta = 0.06$, respectively. We consider $\tau = 0.5$ for all sub-optimal related simulation results. We observe that \gls{NF}-based scheme performs better than all strategies in both the figures, however, the strong user fails in achieving its individual \gls{OMA} rate as is shown in Fig. \ref{fig:MA1} and Fig. \ref{fig:MA6}. Further, it is evident from Fig. \ref{fig:TA1} and Fig. \ref{fig:TA6} that the performance of proposed sub-optimal algorithm is almost equal to the optimal algorithm and is better than the performance of \gls{OMA}. A peculiar remark in the performance metric plot is that the optimal and sub-optimal algorithm's performance always lies between the performance of lower bound and upper bound based power allocations. However, the strong user rate is unable to achieve atleast its \gls{OMA} rate not only in case of \gls{NF}-based pairing, but also in lower bound based power allocation.

The variation of $T_{\alpha}$ with respect to imperfection in \gls{SIC} $\beta$ for $\alpha = 1$ is presented in Fig. \ref{fig:TB1}. We observe that with increase in imperfection in \gls{SIC}, given $\alpha =1$, the parameter metric corresponding to $\alpha$-Fairness scheduler outperforms all other strategies for lower values of $\beta$ and converges with \gls{OMA} for higher values of $\beta$. The performance of \gls{NF}-based pairing strategy decreases beyond \gls{OMA} with increasing $\beta$. This is because, given $\alpha = 1$, we are computing product of strong and weak user rates and with increase in $\beta$, the rate of strong user decreases gradually in \gls{NF} pairing \cite{Mouni}.

Fig. \ref{fig:MA1} and Fig. \ref{fig:MA6} illustrate the mean user rate (MUR) comparison for varying $\alpha$, given the imperfection in \gls{SIC} is $\beta =0.01$ and $\beta =0.06$, respectively. We observe that $\alpha_{ub}$ based power allocation results in better strong user mean rates and the weak user achieves \gls{OMA} equivalent mean user rates, nevertheless, the performance in terms of fairness is poor as observed in Fig. \ref{fig:TA1} and Fig. \ref{fig:TA6}. The $\alpha_{lb}$ based power allocation scheme achieves better weak user mean rates but the strong user mean rates are far below the \gls{OMA} based mean user rates because of the imperfection in \gls{SIC}. Thus, even~though the $T_{\alpha}$ values of $\alpha_{lb}$ based power allocation scheme exceeds the optimal and sub-optimal values as in Fig. \ref{fig:TA1} and Fig. \ref{fig:TA6}, it fails to achieve individual \gls{OMA} rates as the imperfection in \gls{SIC} deteriotes the performance of the strong user. The same is the case with \gls{NF}-based pairing strategy. Even~though \gls{NF} based pairing strategy achieves better $T_{\alpha}$ values in Fig. \ref{fig:TA1} and Fig. \ref{fig:TA6}, it is evident from Fig. \ref{fig:MA1} and Fig. \ref{fig:MA6} that the strong user is unable to achieve atleast \gls{OMA} rates. Lastly, the optimal and sub-optimal algorithm based resultant MURs are always better than their \gls{OMA} counterparts. 


Fig. \ref{fig:MAA1} shows the variation in strong and weak user mean rates for different values of $\beta$ provided $\alpha =1$. The $\delta_{ub}$ based power allocation results in better mean rates than corresponding \gls{OMA} rates. The optimal algorithm based power allocation achieves strong user and weak user mean rates slightly better than \gls{OMA} for lower values of $\beta$. They converge with \gls{OMA} with increasing $\beta$. Though the weak user mean rate in \gls{NF}-based pairing is better than \gls{OMA}, strong user mean rate deteriotes gradually with increasing $\beta$. The $\delta_{ub}$ based power allocation results in strong user mean rates outperforming all other upto a certain $\beta$ and then converges with \gls{OMA}, while the weak user mean rate converges with \gls{OMA} for all values of $\beta$. Lastly, the $\delta_{lb}$ bound weak user mean rate performs slightly better than \gls{OMA} while the strong user mean rate is less than \gls{OMA} for certain values of $\beta$. However, they converge with \gls{OMA} for higher values of $\beta$.

\section{Conclusion}
\label{sec:Conclusion}
We have presented a detailed analysis on $\alpha$-Fairness among the 2-user pair for downlink \gls{NOMA} system in the presence of imperfection in \gls{SIC}. We have formulated the power allocation for the paired users as an optimization problem that achieves $\alpha$-Fairness and ensures minimum of \gls{OMA} rates for the paired users. 
Further, we have also proposed a low-complexity sub-optimal algorithm that achieves close to optimal performance. We have performed extensive simulations and compared the performance of the proposed algorithms against the state-of-the-art algorithms. Both the proposed optimal and the sub-optimal algorithms achieve significant improvements in the network performance. In the future, we plan to implement and evaluate the proposed algorithms in the hardware testbeds.

\bibliographystyle{ieeetran}
\bibliography{Bibfile.bib}

\end{document}